\documentclass{ws-procs9x6}
\newdimen\mathindent
\renewcommand{\qquad}{\hspace*{25pt}}
\newcommand{\eref}[1]{(\ref{#1})}

\def\wPm{$\widehat P$-matrix}

\def\prodd#1#2#3{\prod\limits_{#1}^{#2}\lower3pt\hbox{${ }_{#3}$}}

\newcommand{\eql}[2]{\begin{equation}{#1}\label{#2}\end{equation}}

\def\integer{{\mathchoice
    {\hbox{ $\displaystyle\kern-1mm {\rm Z}\kern-1.1mm {\rm Z}$}}
    {\hbox{ $\textstyle\kern-1mm {\rm Z}\kern-1.1mm {\rm Z}$}}
    {\hbox{$\scriptstyle\kern-1mm {\rm Z}\kern-1.1mm {\rm Z}$}}
    {\hbox{$\scriptscriptstyle\kern-1mm {\rm Z}\kern-1.1mm {\rm Z}$}}}}
\def\natural{{\mathchoice
    {\hbox{ $\displaystyle\kern-1.4mm 1\kern-.7mm {\rm N}$}}
    {\hbox{ $\textstyle\kern-1.4mm 1\kern-.7mm {\rm N}$}}
    {\hbox{$\scriptstyle\kern-1.4mm 1\kern-.7mm {\rm N}$}}
    {\hbox{$\scriptscriptstyle\kern-1.4mm 1\kern-.7mm {\rm N}$}}}}
\def\real{{\mathchoice
    {\hbox{$\displaystyle\kern-.2mm 1\kern-.8mm {\rm R}\kern-.2mm$}}
    {\hbox{$\textstyle\kern-.2mm 1\kern-.8mm {\rm R}\kern-.2mm$}}
    {\hbox{$\scriptstyle\kern-.2mm 1\kern-.8mm {\rm R}\kern-.2mm$}}
    {\hbox{$\scriptscriptstyle \kern-.2mm 1\kern-.8mm {\rm R}\kern-.2mm$}}}}
\def\complex{{\mathchoice
    {\hbox{$\displaystyle\kern-.2mm {\rm C}\kern-1.5mm\raise.2mm
                   \hbox{\vrule height6pt}\kern1.3mm$}}
    {\hbox{$\textstyle\kern-.2mm {\rm C}\kern-1.5mm\raise.3mm
                   \hbox{\vrule height6pt}\kern1.3mm$}}
    {\hbox{$\scriptstyle\kern-.2mm{\rm C}\kern-1.5mm\raise.2mm
                   \hbox{\vrule height3pt}\kern1.3mm$}}
    {\hbox{$\scriptscriptstyle\kern-.2mm{\rm C}\kern-1.5mm\raise.2mm
                  \hbox{\vrule height2pt}\kern1.3mm$}}}}
\def\summ#1#2#3{\sum\limits_{#1}^{#2}\lower3pt\hbox{${ }_{#3}$}}

\def\prodd#1#2#3{\prod\limits_{#1}^{#2}\lower3pt\hbox{${ }_{#3}$}}

\begin{document}

\title{\raggedright
Rational parametrization of strata in orbit spaces of compact
linear groups}

\author{G .SARTORI and G. VALENTE}

\address{Dipartimento di Fisica,Universit\`a di Padova and INFN,
Sezione di Padova\\
I--35131 Padova, Italy \\
E-mail: gianfranco.sartori@pd.infn.it,
valente@pd.infn.it}

%\date{}

\maketitle

%\begin{abstract}
\abstracts{ Functions which are covariant or invariant under the
transformations of a compact linear group $G$ acting in a
euclidean space $\real^n$, can be profitably studied as functions
defined in the orbit space of the group.  The orbit space is the
union of a finite set of strata, which are semialgebraic manifolds
formed by the $G$-orbits with the same symmetry. In this paper we
provide a simple recipe to obtain rational parametrizations of the
strata. Our results can be easily exploited, in many physical
contexts where the study of covariant or invariant functions is
important, for instance in the determination of patterns of
spontaneous symmetry breaking, in the analysis of phase spaces and
structural phase transitions (Landau's theory), in covariant
bifurcation theory, in crystal field theory and in most areas of
solid state theory where use is made of symmetry adapted
functions. An example of utilization of the recipe is also
discussed at the end of the paper.}

%\vskip2truemm PACS: 02.20.Hj, 11.30.Qc, 11.15Ex, 61.50.Ks
%\end{abstract}

\section{Introduction}
The determination of properties of functions which are covariant or invariant under the
transformations of a compact linear group (hereafter abbreviated
in CLG) $G$ is often a basic problem to
solve in many physical contexts. $G$-invariant functions play an important role,
for instance, in the determination of patterns of
spontaneous symmetry breaking and
structural phase transitions (Landau's theory), in covariant
bifurcation theory, in crystal field theory and in most areas of
solid state theory where use is made of symmetry adapted
functions.

It will not be essentially restrictive, in the following, to
assume that $G$ is a matrix subgroup of $O_{n}(\real)$; it will be
thought to act in $\real^n$, thus defining the $G$-space
$(G,\real^n)$. Complex linear groups can always be transformed
into real linear groups through a process of {\em realification} and
compact real linear groups are equivalent to orthogonal linear
groups.

A possible approach to the study of invariant functions, that
fully exploits invariance and regularity properties \cite{020}, is
based on the fact that a $G$-invariant function $f(x)$, $x\in
\real^n$, is a constant along each $G$-orbit and can, therefore,
be considered as a function $\widehat f$ in the orbit space of
$(\real^n/G)$ \footnote{The orbit space $\real^n/G$ is defined as
the quotient space obtained through the equivalence relation
between points belonging to the same orbit), endowed with the
quotient topology and differentiable structure.}.

Geometric invariant theory \cite{611} suggests how to get, in
principle, $\widehat f$ and $\real^n/G$. In fact, theorems by
Hilbert, Schwarz and Nagata \cite{286} assure that a polynomial
or $C^\infty-$, $G$-invariant function $f(x)$, $x\in\real^n$, can
be expressed as a polynomial or, respectively, $C^\infty$ function
$\widehat f(p)$ of a finite minimal set $p(x)=(p_1(x),\dots
,p_n(x))$ of polynomial $G$-invariant functions ({\em minimal
integrity basis} of the ring of $G$-invariant polynomial functions
(hereafter abbreviated in MIB)). The elements of a MIB separate
the orbits of $G$.

Depending on $G$, there may be relations among the elements of a
MIB ({\em non-coregular} $G$):

\eql{\widehat F_A(p(x))\,=\,0,\qquad \forall x\in\real^n\ {\rm
and}\ A=1,\dots,\,.}{e4} The associated equations

\eql{ \widehat F_A(p)\,=\,0,\qquad A=1,\dots, \,,}{e5} define an
algebraic variety in $\real^q$, that is called the {\em variety
$\mathcal Z$ of the relations} (among the elements of the MIB).

The range $p(\real^n)$ of the {\em orbit map} $p$ provides a
diffeomorphic image of $\real^n/G$ in ${\mathcal
Z}\subseteq\real^q$. The set $p(\real^n)$ is a connected
semi-algebraic subset ({\em i.e.}, a subset defined by algebraic
equalities and inequalities) of $\real^q$, formed by the union of
semialgebraic ({\em connected}) manifolds of different dimensions,
called {\em primary strata}. The $G$-orbits with the same isotropy
type form semialgebraic sets ({\em isotropy type strata}), whose
connected components are primary strata. Almost all the points of
$\real^n/G$ belong to a unique stratum, the {\it principal
stratum}, which is a connected open dense subset of $\real^ n/G$.
The boundary of the principal stratum is the union of disjoint
{\it singular} isotropy type strata.  All the strata lying on the
boundary of an isotropy type stratum are open in its topological
closure. The following partial ordering can be introduced in the
set of all the orbit types: $[H]<[K]$, if $H$ is conjugated with a
proper subgroup of $K$. The orbit type of an isotropy type stratum
is contained in the orbit types of all the strata lying in its
boundary; therefore, more peripheral strata are formed by orbits
with higher symmetry under $G$ transformations. The number of
distinct orbit types of $G$ is finite and there is a unique
minimum orbit type, the {\em principal orbit type}, pertaining to
the principal stratum; there is also a unique maximum orbit type
$[G]$, corresponding to the image through $\pi$ of all the points
of $\real^n$, which are invariant under $G$.

The principal stratum is connected, open and dense in $\real^n/G$.
The orbit type $[H]$ of a stratum\footnote{$[H]$ is the class of
subgroups of $G$ conjugate with $H$ and, by definition, is the
class of isotropy subgroups of $G$ at points of the stratum.} is
contained in the orbit types of all the strata lying in its
boundary; therefore, more peripheral strata of $\real^n/G$ are
formed by orbits with higher symmetry under $G$, the {\em
principal orbit type} is minimum and the maximum orbit type is
$[G]$, corresponding to the image through the canonical
projection, $\pi:\ \real^n\mapsto \real^n/G$, of the set of points
of $\real^n$, which are invariant under $G$. This set contains at
least the origin of $\real^n$.

The following theorem offers a simple recipe to derive the
algebraic equations and inequalities defining the range of an
orbit map and its strata \cite{020} (see also \cite{651}).

{\em Let $\widehat P(p)$, $p\in\real^q$ denote the $q\times q$ real symmetric
matrix defined through the relations

\begin{equation}
\widehat P_{ab}(p(x)) = \sum_{j=1}^n\partial_j p_a(x)\,\partial_j
p_b(x),\quad \forall x\in {\bf R}^n.\label{P}
\end{equation}
Then, $p(\real^n)=\{p\in{\mathcal Z} \mid \widehat P(p)\ge 0\}$ and
the $k$-dimensional primary strata are the connected components of
the semialgebraic set $\{p\in{\mathcal Z}\mid \widehat P(p)\ge 0,\ {\rm
rank}(\widehat P(p))=k\}$.}

If $G$ is coregular, the $p_i$'s, with range in the semialgebraic
set $\Delta$ defined by the inequalities $\widehat P(p)>0$ (and no
equation!) provide a one-to-one parametrization of the principal
stratum of $\real^n/G$. If the stratum is singular, the $p_i$'s
are not independent parameters, being bounded also by the
equations defining the stratum. In applications or, simply, to get
a better understanding of the geometry of the stratum (its
connectivity properties, its boundary, etc..), one often needs to
solve explicitly these equations, which are obtained in implicit
form from the proposition recalled above. Often, this cannot be
done using standard algorithms.

In this paper we propose a general method to derive the equations
defining the singular strata of the orbit space of any CLG in the
form of explicit rational parametric relations. In the next
Section we shall present and prove our results and in Section 3,
we shall illustrate them, by working out an example.

\section{Parametrization of strata in orbit spaces of compact linear groups}

In this section, we shall prove that the possibility of
parametrizing the principal stratum of the orbit space of a
coregular group in terms of parameters related to the elements of
a MIB can be extended to singular strata. The result stems from
the proof of the following statement: {\em The set of internal
points of the topological closure of an isotropy type stratum,
with orbit type $[H]$, is diffeomorphic to the principal stratum
of the orbit space of a group space of the stabilizer of $H$ (in
$G$).} If the stabilizer turns out to be coregular, the
constructive proof of this proposition will provide a one-to-one
rational parametrization of the stratum.

Since, in view of possible applications, we are mainly interested
in the possibility of getting a one-to-one parametrization of a
stratum, in the presentation of our results, stress will be laid
on this aspect of the matter.

Let $G$ and $\{p\}$ be defined as in the previous section (no
coregularity assumption will be done on $G$) and $H$ be a proper
isotropy subgroup of the $G$-space $(G,\real^n)$. We shall denote
by $S$ the stratum of $(G,\real^n)$ with orbit type $[H]$ and by
$\widehat S$ its image in the orbit space $\real^n/G$. In the following, we shall
identify $\real^n/G$ and its strata $\widehat S$ with their images $p(\real^n)$ and
$p(\widehat S)$. Let us, also, define

\begin{equation}
\begin{array}{rcl}
V &=& \{x\in\real^n\mid G_x\supseteq H\},\\
&&\\
{\mathcal V}&=&\{x\in\real^n \mid G_x = H\}\,.
\end{array}
\label{1}
\end{equation}

The non trivial sets $V$  and ${\mathcal V}$ have the following properties, which are more or
less immediate consequences of their definitions:

\begin{enumerate}
\item $V =\{x\in\real^n \mid h x=x, h\in H\}$ is a linear subspace of $\real^n$, let us call
$\nu$ its dimensions;
\item $V$ is the topological closure of $\mathcal V$: $V=\overline{\mathcal V}$;
\item ${\mathcal V}=S \cap V $, so that every $G$-orbit lying in $S$ has at least a point in
${\mathcal V}$ and $S = G\cdot {\mathcal V}$, $\overline{S}=G\cdot V$, where the bar denotes
topological closure, so that

\eql{p({\mathcal V})={\widehat S},\qquad p(V)=\overline{\widehat
S}}{pV}
and

\eql{{\rm rank}(P(x)\big|_{x\in{\mathcal V}})={\rm dim}(\widehat S),\qquad
{\rm rank}(P(x)\big|_{x\in V\backslash{\mathcal V}})\,<\,{\rm dim}(\widehat S).}{rango}
\end{enumerate}

A $G$-orbit of $S$ may intersect $\mathcal V$ in one or more
points. In the first case, every $G$-orbit of $S$ intersects
$\mathcal V$ in a point and, owing to item 3, the coordinates of
the points of ${\mathcal V}$ provide a one-to-one rational (in
effect, linear) parametrization of the orbits of G lying in $S$,
obtained by solving the system of linear equations $h\,x=x\
\forall h\in H$. The allowed range of $x$, $x\in{\mathcal V}$, is
determined by the inequalities assuring that
rank$(P(x))=$\,dim$(\widehat S)$. The parameters $x$, with range
$\mathcal V$, would provide, in this case, the one-to-one
parametrization of the stratum, thus solving our problem.

In general, however, the intersection of a $G$-orbit, of orbit type $[H]$,
with the stratum $S$ does not reduce to a single point. So, a sounder
analisys is required, that will be the object of the rest of this section.

Two distinct points, $x$ and $g\,x$, $g\in G$, of the same $G$-orbit, lay in
$\mathcal V$ iff $h\, g\, x = g\, x\ \forall h\in H$. This is
equivalent to the condition that $g$ is an element of the stabilizer Stab$(H, G)$ of
$H$ in $G$:

\eql{ {\rm Stab}(H, G)=\{g\in G\mid g H g^{-1} = H\}.}{0}

The intersection of a $G$-orbit of $S$ with ${\mathcal V}$ is, therefore,
the Stab($H$, $G$)-orbit through $x$.

This result can be refined. In fact, it is easy to realize that
the relation $h\, g\, x\, = g\, x$ is satisfied for all $x\in V$
and all $h\in H$, iff $g\in$ Stab($H$, $G$). Therefore, Stab($H$,
$G$) {\em is the largest subgroup of $G$ under which $V$ is
stable}; a known result.

In the group space (Stab($H$, $G$),V), the isotropy subgroup at a point
of general position is $H$ (or, more precisely, the linear group
obtained by restriction to $V$ of the action of $H$). Therefore,
the principal isotropy type is $[H]=\{H\}$ and the principal stratum $\Sigma$
satisfies the following relation:

\eql{V\supset \Sigma\supseteq {\mathcal V}.}{ins}
It has to be noted that, the principal stratum $\Sigma$ could contain
$\mathcal V$ in a strict sense, since, at points of $V\backslash{\mathcal V}$,
the conjugacy class of the isotropy subgroup of Stab$(H,G)$ could be
smaller than the conjugacy class of the isotropy subgroup of $G$.

$H$ is an invariant
subgroup of Stab$(H,G)$ and it is a subgroup of all the isotropy
subgroups of (Stab$(H,G),V)$. Thus, the restriction of the action
of Stab$(H,G)$ to $V$ defines a linear group $K$ (and the group space
$(K,V)$), isomorphic to the quotient group Stab($H$, $G$)/$H$, through the relation

\eql{(sH)\,v=s\,v,\qquad s\in {\rm Stab}(H,G),\ sH\in {\rm
Stab}(H,G)/H.}{K}
So, we can conclude that {\em the intersection
of a $G$-orbit of $S$ with ${\mathcal V}$ is an orbit of $(K,V)$}. Obviously,
the orbit spaces $V/K$ and $V/{\rm Stab}(H,G)$ are isomorphic and can be
identified.

We can rephrase the result just obtained, by claiming that the
points of $\widehat S$ are in a one-to-one correspondence with the
points of a, possibly proper, subset $\widehat{\mathcal V}$ of the principal stratum
$\widehat\Sigma$ of the orbit space $V/K$ and the following relations hold true:

\eql{\overline{\widehat S}\,=\,p(V)\,\supset\, p(\Sigma)\,\supseteq\, p({\mathcal V})\, =\, \widehat S.}{pS}
Since, as stressed, we know how to parametrize
a principal stratum, our problem is reduced to the determination of $\widehat{\mathcal V}$.
We shall show that this can be easily done, making use of \eref{rango}.

To make simpler the solution of the problem, let us assume that an orthonormal basis
has been introduced in $\real^n$ such that the first $\nu$ elements of the basis yield
a basis for the vector space $V$. Then, if we denote by $V_\bot$ the orthogonal complement
of $V$ in $\real^n$, the subspace $V_\bot$ is stable under ($K$ and) $H$, owing to the
orthogonality of the transformations of $G$. Since $V$ contains all the $H$-invariant vectors of
$\real^n$, there is no non trivial $H$-invariant vector in $V_\bot$.

To attain our goal, we shall start from a convenient
parametrization of the principal statum $\widehat \Sigma$ of $V/K$
in terms of $l$ real parameters $\lambda$, related to a MIB
$(\lambda_1(v),\dots ,\lambda_l(v))$ of the ring of polynomial
$K$-invariant functions. This parametrization will be global and
one-to-one, if $K$ turns out to be coregular. In this case, the
range of $\lambda$ has to be restricted to the positivity region
of the $\widehat P$-matrix $\widehat\Lambda(\lambda)$, associated
to the MIB $\{\lambda\}$:

\eql{\widehat\Lambda_{\alpha\beta}(\lambda(v)) = \Lambda_{\alpha\beta}(v) = \sum_{i=1}^\nu\,
\partial_i \lambda_\alpha(x_1,\dots ,x_\nu)\, \partial_i \lambda_\beta(x_1,\dots ,x_\nu).}{4}
As recalled in the previous Section, in fact, the orbit
space $V/K$ and its principal stratum $\widehat \Sigma$ can be
identified, respectively, by the semialgebraic sets $\lambda(V)$ and
$\lambda(\Sigma)$:

\eql{\widehat\Sigma \,=\, \lambda(\Sigma)\,=\,\{\lambda\in\real^l\mid \widehat
\Lambda(\lambda)>0\},\qquad
\overline{\widehat \Sigma}\,=\,\lambda(V)\,=\,\{\lambda\in\real^l\mid
\widehat\Lambda(\lambda)\ge 0\},}{20}
the second set being the closure of the first one. Moreover, the definition
of $\widehat{\mathcal V}$ and equation \eref{ins} imply

\eql{\widehat{\mathcal V}=\lambda({\mathcal V})\subseteq\lambda(\Sigma)\,=\,\widehat\Sigma.}{inshat}

If $K$ is not coregular, only a local one-to-one parametrizations
can be obtained for $\widehat \Sigma$, by eliminating redundant
elements in the set of parameters $(\lambda_1,\dots ,\lambda_l)$,
through the solution of the algebraic relation(s) among the
elements of the MIB $\{\lambda\}$, and imposing convenient
semi-positivity and rank conditions on the matrix
$\widehat\Lambda(\lambda)$.

A one-to-one local or global parametrization of $\widehat \Sigma$
yields, obviously, also a local or global one-to-one
parametrization of $\widehat {\mathcal V}$, provided that
additional restrictions are imposed on the range of $\lambda$,
whenever $\widehat {\mathcal V}$ is a proper subset of
$\widehat\Sigma$. The correct bounds can be obtained in the
following way.

We shall only consider the case of a coregular $K$, the extension
of the results to non coregular $K$'s will be straightforward,
but, as just stressed, may lead to a loss of globality.

When $x$ spans $V$, the elements of the MIB
$(p_1(x),\dots,p_q(x))$ of $G$ define a set of $K$-invariant
polynomial functions of $v=(x_1,\dots ,x_\nu)\in V$. Therefore, by
the Hilbert theorem recalled in the previous Section, they can be
expressed as polynomial functions of $\lambda$, that is,

\eql{p\big|_V=\phi\circ\lambda.}{phi}

Let us remark that, in our assumptions, there are no relations among the elements of the MIB
$\{\lambda\}$ and, consequently, possible relations $F_\alpha(p)=0$ among the $p_i$'s are
identically satisfied for $p=\phi(\lambda)$.

>From \eref{phi} and \eref{ins}, one immediately obtains

\eql{\overline{\widehat S}\,=\,p(V)\,=\,\phi(V/K)\,\supset\,
\phi(\widehat\Sigma)\,\supseteq\,\phi(\widehat{\mathcal V})\,=\,p({\mathcal V})=\widehat S.}{pins}
and, since $\widehat\Sigma$ is the set of internal points of $V/K$,
$\phi(\widehat\Sigma)$ will coincide with the set of
internal points of the closure $\overline{\widehat S}$ of $\widehat S$. This set
does not coincide with $\widehat S$ if
$\overline{\widehat S}$ contains, in its interior, points representing
bordering strata of $\widehat S$. This happens if the set
$\widehat S$ is not connected or not multiply connected.

The identification of points $\lambda\in\widehat \Sigma$, whose image
$\phi(\lambda)\not\in\widehat S$ can be obtained in the following way,
using \eref{rango}.

Let $x=v\oplus v_\bot$ yield the decomposition of $x\in\real^n$
in its vector components $v\in V$ and $v_\bot\in V_\bot$:
$v=(x_1,\dots ,x_\nu)$, $v_\bot=(x_{\nu+1},\dots ,x_n)$. Then,
a $G$-invariant polynomial $f(x)$, can be thought of as a polynomial
in $v$ and $v_\bot$ and it is easy to realize that it cannot contain linear terms in $v_\bot$,
being $v$ invariant under $H$. As a consequence,

\eql{\partial_i f(x)=0,\ {\rm for\ }x\in V\ {\rm and\ }i=\nu+1,\dots ,n.}{3}

Now, starting from the very definition of $P(x)$ (see \eref{P}),
for every $x=v\in V$ we obtain, using \eref{3} and the identity
$p(v)=\phi(\lambda(v))$,

\eql{\begin{array}{rcl}
\widehat P_{ab}(p(v))&=&\sum_{i=1}^\nu\,\partial_i p_a(v)\,\partial_i p_b(v)\\
&&\\
&=& \left. \sum_{\alpha,\beta=1}^l\,\partial_\alpha\phi_a(\lambda)\,
\partial_\beta\phi_b(\lambda)
\right|_{\lambda=\lambda(v)}\,\sum_{i=1}^\nu\,\partial_i \lambda_\alpha(v)\,
\partial_i\lambda_\beta(v)\\
&&\\
&=& \left. \left(J(\lambda)^T\,\widehat\Lambda(\lambda)\,J(\lambda)\right|_{\lambda=\lambda(v)}\right)_{ab}
,
\end{array}}{5}
where, the apex $T$ denotes transposition and $J(\lambda)$ is the Jacobian matrix of
the transformation $p=\phi(\lambda)$:

\eql{J_{\alpha a}(\lambda)=\partial_a\phi_\alpha (\lambda),\qquad
a=1,\dots ,q,\ \ \alpha=1,\dots ,l.}{7}
So, we can conclude that, for all $\lambda\in V/K$,

\eql{\widehat P(\phi(\lambda))= J(\lambda)^{\rm T}\,\widehat\Lambda(\lambda)\,J(\lambda).}{6}

Equation \eref{6} leads to an easy identification of
the points $\lambda\in \widehat{\mathcal V}$, that is, of the points $\lambda\in \widehat\Sigma$ whose
image $p=\phi(\lambda)$ is in $\widehat S$. In fact,
from \eref{rango}, these points are characterized by the conditions $\lambda\in\widehat\Sigma$ and
rank($\widehat P(\phi(\lambda)))=l$. Since, for $\lambda\in\widehat\Sigma$,  the matrix
$\widehat \Lambda(\lambda)$ is positive definite with rank $l$,
we can conclude that the range of $\lambda$ assuring the location of the point
$p=\phi(\lambda)$ in $\widehat S$ is the semialgebraic subset $\Delta$ of $\real^l$,
determined by the following inequalities:

\eql{\Delta=\{\lambda\in\real^l\mid \Lambda(\lambda)>0 \ {\rm
and}\ {\rm rank}(J(\lambda))=l\}.}{8} For $\lambda\in\Delta$, the
relation $p=\phi(\lambda)$ yields a global rational
parametrization for $\widehat S$.

It will be worthwhile to stress that the parametrization we have
suggested is, in some way, canonical: The unique arbitrariness in
the choice of the parameters is related to the choice of the
MIB's.

An important byproduct of the result just proved is a simple test of the connection of $\widehat S$,
that could be difficult or impossible to read directly from the equations of the stratum in implicit form.
In fact, the condition

$${\rm boundary\ of\ }\overline{\widehat S}\,=\,{\rm  boundary\ of\ }\widehat S$$
is equivalent to rank($J(\lambda))=l$ for all $\lambda\in\widehat \Sigma$.

If $K$ is not coregular, there are polynomial relations
$F_A(\lambda)=0$ among the elements of the MIB $(\lambda_1,\dots
,\lambda_l)$ and $l>$dim$(\widehat S)$. As already stressed, to
obtain a one-to-one parametrization of the points of $\widehat S$
by means of the $\lambda_\alpha$'s, one has to eliminate the
redundant parameters by solving the equations $F_A(\lambda)=0$.
This may be feasible, but only locally and the resulting
parametrization will not be global and possibly not rational.

\section{An Example}
In this Section we shall show how the parametrization technique
works in a simple example. The notations will be the same defined
in the previous sections.

We shall consider the orthogonal linear group $G$ defined by the action of the real group $O(3)$ in the real
eighth-dimensional space spanned by the independent components $x_1,\dots ,x_5$ of a symmetric
and traceless tensor $Q$ and the three components $x_6,x_7,x_8$ of a polar vector $\vec{P}$.
To be specific, if $O$ is a generic $3\times 3$ real orthogonal matrix,
the transformation rules of the $x_i$ are obtained from the following relations:

\begin{equation}
Q = \frac{1}{\sqrt{2}} \left(
\begin{array}{ccc}
-{\displaystyle\frac{2}{\sqrt{3}}} x_1&  x_3 & x_4 \\
x_3                       & {\displaystyle\frac{x_1}{\sqrt{3}}} - x_2 & x_5 \\
x_4                       &  x_5                & {\displaystyle\frac{x_1}{\sqrt{3}}} + x_2 \\
\end{array}
\right)
\end{equation}

\begin{equation}
\begin{array}{rcll}
Q'_{\alpha\beta}&=& \sum_{\gamma\,\delta=1}^3 O_{\alpha\gamma}\,
                                    Q_{\gamma\delta}\,  O_{\beta\delta}& \\
                                    &&&\alpha,\beta=1,2,3\,.    \\
P'_{\alpha}&=& \sum_{\beta=1}^3 O_{\alpha\beta}\, P_{\beta}\;,
\end{array}\label{mu2}
\end{equation}
It will be worthwhile to remark that the reflection $I=\mbox{\rm diag}
(-1,-1,-1)\in$ O(3) reverses only
the signs of the last three coordinates $(x_6,x_7,x_8)$ and, due to the
symmetry of the tensor $Q$, each $G$-orbit contains points where
the tensor $Q$ takes on a diagonal form ($x_3=x_4=x_5=0)$. This remark makes much easier the
determination of ``typical points" in strata, from which the orbit type of the stratum can be
identified.

The real orthogonal linear group $G$, just defined, is coregular (see \cite{710}) and its
isotropy subgroup at a generic point of $\real^8$ is trivial.
As a consequence, the principal orbits are 3-dimensional manifolds, the
principal stratum has dimensions 5 and there will be five independent elements in a MIB
(see, for instance \cite{683}), which can be chosen in the following way:

\begin{eqnarray}
p_1 &=& \mbox{\rm Tr}\, Q^{2} + \overrightarrow{P}\cdot
\overrightarrow{P} \nonumber\\
    &=& \sum_{j=1}^{8} {x_j}^{2}\,,\nonumber\\
&&\nonumber\\
p_2 &=& \overrightarrow{P}\cdot \overrightarrow{P} \nonumber\\
    &=&  \sum_{j=6}^{8} {x_j}^{2}\,,\nonumber\\
&&\nonumber\\
p_3 &=& 6 \sqrt{2} \,\,\mbox{\rm Tr}\, Q^{3}\nonumber\\
    &=& -2\, {\sqrt{3}}\, {{x_1}}^3 + 6\, {\sqrt{3}}\, {x_1}\, {{x_2}}^2
    - 3\, {\sqrt{3}}\, {x_1}\, {{x_3}}^2 - 9\,  {x_2}\, {{x_3}}^2
      \label{BaseInt}\\
  && - 3\, {\sqrt{3}}\, {x_1}\, {{x_4}}^2+ 9\, {x_2}\, {{x_4}}^2+18\,
{x_3}\, {x_4}\, {x_5} +   6\, {\sqrt{3}}\, {x_1}\, {{x_5}}^2 \,,\nonumber\\
&& \nonumber\\
p_4 &=& 3 \sqrt{2} \,\sum_{\alpha\,\beta}P_\alpha Q_{\alpha\,\beta}
P_{\beta}\nonumber\\
 &=& -2\,{\sqrt{3}}\, {x_1}\, {{x_6}}^2  +
  6\, {x_3}\, {x_6}\, {x_7} + {\sqrt{3}}\, {x_1}\, {{x_7}}^2 -
  3\, {x_2}\, {{x_7}}^2  \nonumber\\
  && + 6\, {x_4}\, {x_6}\, {x_8}+ 6\, {x_5}\, {x_7}\, {x_8}+ \sqrt{3}\,
{x_1}\, {{x_8}}^2+3\, {x_2}\, {{x_8}}^2\,,\nonumber\\
&&\nonumber\\
p_5 &=& 6 \,\sum_{\alpha\,\beta}P_\alpha {Q^{2}}_{\alpha\,\beta}
P_{\beta}\nonumber\\
    &=& 4\,{{x_1}}^2\, {{x_6}}^2+ 3\, {{x_3}}^2\, {{x_6}}^2 +
  3\, {{x_4}}^2\, {{x_6}}^2-2\, {\sqrt{3}}\, {x_1}\, {x_3}\,
{x_6}\,{x_7}
    \nonumber\\
  &&- 6 \,{x_2}\,{x_3}\,{x_6}\,{x_7}+ 6 \,{x_4}\,{x_5}\,{x_6}\,{x_7} +
   {{x_1}}^2\, {{x_7}}^2 -
  2\, {\sqrt{3}}\, {x_1}\, {x_2}\, {{x_7}}^2 \nonumber\\
  &&+ 3\, {{x_2}}^2\, {{x_7}}^2 +
  3\, {{x_3}}^2\, {{x_7}}^2 + 3\, {{x_5}}^2\, {{x_7}}^2 -
  2\, {\sqrt{3}}\, {x_1}\, {x_4}\, {x_6}\, {x_8}   \nonumber\\
  && +  6\, {x_2}\, {x_4}\, {x_6}\, {x_8}+6\, {x_3}\, {x_5}\, {x_6}\,
  {x_8} +     6\, {x_3}\, {x_4}\, {x_7}\, {x_8} +
  4\, {\sqrt{3}}\, {x_1}\, {x_5}\, {x_7}\, {x_8}   \nonumber\\
  &&+   {{x_1}}^2\, {{x_8}}^2
  +2\,{\sqrt{3}}\, {x_1}\, {x_2}\, {{x_8}^2}+3\, {{x_2}}^2\, {{x_8}}^2 +
  3\, {{x_4}}^2\, {{x_8}}^2   +3\, {{x_5}}^2\, {{x_8}}^2\;.\nonumber
  \end{eqnarray}

The corresponding $\widehat P $-matrix elements can be easily calculated
from their very definition \eref{P}

\begin{equation}
\begin{array}{rcl}
\widehat{P}_{1\,a} &=& 2 d_a p_a\;, \hspace{3em} 1 \leq a \leq 5 \\
\widehat{P}_{2\,2} &=& 4 p_2 \\
\widehat{P}_{2\,3} &=& 0 \\
\widehat{P}_{2\,4} &=& 4 p_4 \\
\widehat{P}_{2\,5} &=& 4 p_5 \\
\widehat{P}_{3\,3} &=& 108 \left(p_1 -p_2\right)^{2} \\
\widehat{P}_{3\,4} &=& 18\,\left( - 2\,p_1\,p_2+ 2 {p_2}^2+ p_5 \right) \\
\widehat{P}_{3\,5} &=& 12\,\left( \,p_2\,p_3+ p_1 \,p_4 -p_2 \, p_4 \right) \\
\widehat{P}_{4\,4} &=& 12 \,\left({p_2}^2 + p_5 \right)\\
\widehat{P}_{4\,5} &=& 4\,\left( \,p_2\,p_3+3\, p_1 \,p_4 -p_2 \, p_4 \right) \\
\widehat{P}_{5\,5} &=& \frac{4}{3} \left(p_3 \, p_4 + {p_4}^2 + 9\, p_1 p_5 \right)\;,
\end{array}
\end{equation}
where the set $\{d_1,d_2,d_3,d_4,d_5\}=\{2,2,3,3,4\}$ specifies the degrees of the polynomials
of the MIB.

The determinant of the matrix $\widehat P(p)$ factorizes and
only one of the two real irreducible factors,
that we shall call $A(p)$, turns out to be {\em active} \cite{682}, that is, to be
relevant in the determination of the boundary of the orbit space $\real^8/G$:

\begin{equation} \label{attivo}
\begin{array}{rcl}
A(p)&=& 3\,{{p_2}}^3\,{{p_3}}^2 + 18\,{p_1}\,{{p_2}}^2\,{p_3}\,{p_4} -
  18\,{{p_2}}^3\,{p_3}\,{p_4} +
  27\,{{p_1}}^2\,{p_2}\,{{p_4}}^2 \\
  && -
  54\,{p_1}\,{{p_2}}^2\,{{p_4}}^2 + 27\,{{p_2}}^3\,{{p_4}}^2 +
  {p_3}\,{{p_4}}^3 - 9\,{p_2}\,{p_3}\,{p_4}\,{p_5}  \\
  &&
  -9\,{p_1}\,{{p_4}}^2\,{p_5} + 9\,{p_2}\,{{p_4}}^2\,{p_5} -
  27\,{p_1}\,{p_2}\,{{p_5}}^2 + 27\,{{p_2}}^2\,{{p_5}}^2 +
  9\,{{p_5}}^3\;.\\
\end{array}
\end{equation}

The relations assuring that $\widehat P(p)\ge 0$
and has rank 4, define a unique 4-dimensional stratum $\widehat S^{(4)}$ in the orbit space. Using
well known matrix theory results, these conditions could be explicitly written, for instance,
in the form $A(p)=0$ and $M_i(p)>0$, $i=1,\dots ,4$,
where $M_i$ is the sum of the order $i$ principal minors of the matrix $\widehat P(p)$: A
frightening set of conditions that it is not worthwhile to write down explicitly.

Even in this simple example one immediately realizes that the
choice of a minimal set of explicit algebraic relations providing
a cylindrical decomposition for the semi-algebraic subset
$\widehat S^{(4)}$ of $\real^{5}$ would be a really hard task (for
the more peripheral strata, instead, the problem is much easier to
solve). An immediate application of the results proved in the
previous Section, on the contrary, leads to a simple rational
global parametrization of the sub-principal stratum $\widehat
S^{(4)}$.

A ``typical point" of
the stratum is $x_{\rm t}=(1,1,0,0,0,0,1,1)$. The corresponding isotropy subgroup
$H$ of $G$ is the $\integer_2$ group generated by the reflection

$$\mbox{\rm diag}(1,1,-1,-1,1,-1,1,1)\in G,$$
representing the element diag$(-1,1,1)\in$ O(3).

The vector space $V$ formed by the $H$-invariant vectors of $\real^n$ turns out to
be be the following:

$$V = \left\{x\in \real^8 \;|\;x_3=x_4=x_6=0\right\}.$$

The elements
of  O(3), corresponding to elements of $\mbox{\rm Stab }(H,G)$, are block-diagonal matrices
of the form diag$(\pm 1,O)$, with $O\in $ O(2). Therefore, $\mbox{\rm Stab }(H,G)$ is isomorphic
to a group $\integer_2\otimes$O(2) and the quotient group $K=\mbox{\rm Stab }(H,G)/H$ is the
representation in $\real^8$ (induced by the representation $G$ of O(3)) of the
O$_1(2)$ subgroup of O(3), formed by the rotations around the first axis.

Using coordinate $v=(x_1, x_2,x_5,x_7,x_8)$ for a vector $v\in V$, the elements of $K$
turn out to be block diagonal matrices leaving stable the following subspaces
$V^{A}$, $1 \leq A \leq 3$:

\begin{eqnarray}
V^{1}& = &\{(x_1,0,0,0,0) \in V | x_1 \in \real\}\,, \nonumber \\
V^{2}& = &\{(0,x_2,x_5,0,0) \in V | x_2\,,x_5  \in \real\} \,,\nonumber \\
V^{3}& = &\{(0,0,0,x_7,x_8) \in V | x_7,,x_8 \in \real\}\,.
\end{eqnarray}

A proper rotation $r(\phi)\in$ O$_1(2)$ of an angle $\phi$ and the reflection diag(-1,1)$\in {\rm O}_1(2)$
are represented in the primed basis of $V$ by the $5\times 5$ matrices
$h(\phi)=1\oplus r(-2\phi)\oplus r(\phi)$ and $j={\rm diag}(1,-1,1,-1,1)$.

Noting that the complex variables $z_1=x_2+i x_5$ and  $z_2=x_7+i x_8$ transform
into $e^{-2i \phi}z_1$ and, respectively,  $e^{i \phi}z_2$ under a rotation $\tilde h(\phi)$ and
into $z_1^*$ and, respectively, $-z_2^*$ under a reflection $j$, it is easy to realize that a possible
MIB. for $(\widetilde{H}, V)$ is the following:

\begin{eqnarray}
\lambda_1 &=& x_1 \,,\nonumber\\
\lambda_2 &=&|z_1|^2\,=\, x_2^2 + x_5^2\,,\nonumber\\
\lambda_3 &=& |z_2|^2\,=\,x_7^2 + x_8^2\,,\nonumber\\
\lambda_4 &=& 2 \, \mbox{\rm Re} (z_1 {z_2}^2)\,=\,2
\left( x_2 {x_7}^2 -2 x_5 x_7 x_8 - x_2 {x_8}^2 \right)\;.
\end{eqnarray}

It is, now, easy to express the $p$'s in term of the $\lambda$'s, $p=\phi(\lambda)$, and to check that
the following expressions, obtained in this way, identically satisfy the equation
$A(p(\lambda))=0$ (see (\ref{attivo})) entering the definition of the stratum $\widehat S^{(4)}$:

\begin{eqnarray} \label{parametrizzazione}
\phi_1(\lambda) &=& {\lambda_1}^2 + \lambda_2 + \lambda_3 \,,\nonumber\\
\phi_2(\lambda) &=& \lambda_3\,,\nonumber\\
\phi_3(\lambda) &=& 2 \, \sqrt{3}\,\left( {\lambda_1}^3 - 3 \lambda_1
\lambda_2\right)\,,\nonumber\\
\phi_4(\lambda) &=& - \frac{\sqrt{3}}{2} \,\left( 2 \lambda_1 \lambda_3 + \sqrt{3}
\lambda_4 \right)\,, \nonumber\\
\phi_5(\lambda) &=& {\lambda_1}^{2} \lambda_3 + 3 \lambda_2 \lambda_3 +
\sqrt{3} \lambda_1 \lambda_4\,.
\end{eqnarray}

As explained in the previous section, since the group $K$ is coregular,
the range  $\Delta$ for $\lambda$, is
the region where the \wPm, $\widehat{\Lambda}(\lambda)$ associated to the MIB $\{\lambda\}$ is
positive definite and the rank of the jacobian matrix $J(\lambda)$ of the transformation
$\phi(\lambda)$ is maximum (=4). The matrices $\widehat\Lambda(\lambda)$ and $J(\lambda)$
are easily calculated to be

\begin{equation} \label{lamdilam}
\widehat{\Lambda}(\lambda) = \left(
\begin{array}{cccc}
1 & 0 & 0 & 0 \\
0 & 4 \lambda_2 & 0 & 2 \lambda_4 \\
0 & 0 & 4 \lambda_3 & 4 \lambda_4 \\
0 & 2 \lambda_4 & 4 \lambda_4 & 4\,\left( {\lambda_3}^2 + 4
\lambda_2 \lambda_3 \right)
\end{array}
\right)
\end{equation}
and

\begin{equation}
J(\lambda)^{\rm T} = \left(
\begin{array}{cccc}
2 \lambda_1 & 1 & 1 & 0 \\
0 & 0 & 1 & 0 \\
6\,\sqrt{3} \,\left({\lambda_1}^2 - \lambda_2 \right) &
        - 6\,\sqrt{3}\, \lambda_1 & 0 & 0 \\
- \sqrt{3} \lambda_3 & 0 & - \sqrt{3} \lambda_1 & -\frac{3}{2} \\
2 \lambda_1 \lambda_3 + \sqrt{3} \lambda_4 & 3 \lambda_3 &
{\lambda_1}^2 + 3 \lambda_2 & \sqrt{3} \lambda_1
\end{array}
\right)\;.
\end{equation}
The conditions assuring the positivity of $\widehat{\Lambda}(\lambda)>0$ can be written
in the form

\begin{equation}
\Delta = \left \{(\lambda_1,\lambda_2, \lambda_3, \lambda_4) \in
\real^4 \;|\; \lambda_2 >0 \; \mbox{and} \; \lambda_3 >0
\;\mbox{and}\; \lambda_4^2 < 4 \lambda_2 {\lambda_3}^2 \right \}
\end{equation}
and the rank of $J(\lambda)$ turns out to equal four for all $\lambda \in \Delta$.
So we can conclude that the stratum is connected, a piece of information that would be
difficult to derive directly from the relations defining the stratum in implicit form
($(A(p)=0$ and $M_i(p)>0$).

The parametrization obtained for the sub-peripheral stratum
$\widehat S^{(4)})$ turns out to be useful, also, because
bordering values of $\lambda$ immediately determine bordering
values of $p$, corresponding to more peripheral strata of
$\real^8/G$.
%The stratification of $ V/K$ is summarized in Table~(\ref{tab2}).

%\begin{table}
%\caption{ \label{tab2}
%Isotropy type stratification of the orbit space $V/K$, whose principal stratum is
%diffeomorphic to the subprincipal stratum $\widehat S^{(4)}$ of the orbit space $\real^8/G$.}
%\begin{center}
%\begin{tabular}{|c|c|c|}
%\hline
%&&\\
%Stratum & Equalities & Inequalities \\
%&&\\
%\hline
%&&\\
%$\widehat\Sigma^{(2)}$ & $\lambda_3=\lambda_4= 0$ & $\lambda_2>0$ \\
%&&\\
%\hline
%&&\\
%$\widehat\Sigma^{(3)}$ & $ 4 \lambda_2  {\lambda_3}^{2}= {\lambda_4}^2$ & ${\lambda_3}>0$
%and
%$\lambda_2 \geq 0$ \\
%&&\\
%\hline
%&&\\
%$\widehat\Sigma_p$ &    & $\lambda_2>0$ and $\lambda_3>0$ and
%$ 4 \lambda_2  {\lambda_3}^{2}- {\lambda_4}^2 >0$ \\
%&&\\
%\hline
%\end{tabular}
%\end{center}
%\end{table}

\section*{Aknowledgements}
This paper is partially supported by INFN and MIURST.

\end{document}